\documentstyle[twoside]{article}

\catcode`\@=11
\long\def\@makefntext#1{
\protect\noindent \hbox to 3.2pt {\hskip-.9pt  
$^{{\eightrm\@thefnmark}}$\hfil}#1\hfill}		

\def\thefootnote{\fnsymbol{footnote}}
\def\@makefnmark{\hbox to 0pt{$^{\@thefnmark}$\hss}}	
	
\def\ps@myheadings{\let\@mkboth\@gobbletwo
\def\@oddhead{\hbox{}
\rightmark\hfil\eightrm\thepage}   
\def\@oddfoot{}\def\@evenhead{\eightrm\thepage\hfil
\leftmark\hbox{}}\def\@evenfoot{}
\def\sectionmark##1{}\def\subsectionmark##1{}}

\oddsidemargin=\evensidemargin
\renewcommand{\thefootnote}{\fnsymbol{footnote}}

\newcounter{sectionc}\newcounter{subsectionc}\newcounter{subsubsectionc}
\renewcommand{\section}[1] {\vspace{12pt}\addtocounter{sectionc}{1} 
\setcounter{subsectionc}{0}\setcounter{subsubsectionc}{0}\noindent 
	{\tenbf\thesectionc. #1}\par\vspace{5pt}}
\renewcommand{\subsection}[1] {\vspace{12pt}\addtocounter{subsectionc}{1} 
	\setcounter{subsubsectionc}{0}\noindent 
	{\bf\thesectionc.\thesubsectionc. {\kern1pt \bfit #1}}\par\vspace{5pt}}
\renewcommand{\subsubsection}[1] {\vspace{12pt}\addtocounter{subsubsectionc}{1}
	\noindent{\tenrm\thesectionc.\thesubsectionc.\thesubsubsectionc.
	{\kern1pt \tenit #1}}\par\vspace{5pt}}
\newcommand{\nonumsection}[1] {\vspace{12pt}\noindent{\tenbf #1}
	\par\vspace{5pt}}

\newcounter{appendixc}
\newcounter{subappendixc}[appendixc]
\newcounter{subsubappendixc}[subappendixc]
\renewcommand{\thesubappendixc}{\Alph{appendixc}.\arabic{subappendixc}}
\renewcommand{\thesubsubappendixc}
	{\Alph{appendixc}.\arabic{subappendixc}.\arabic{subsubappendixc}}

\renewcommand{\appendix}[1] {\vspace{12pt}
        \refstepcounter{appendixc}
        \setcounter{figure}{0}
        \setcounter{table}{0}
        \setcounter{lemma}{0}
        \setcounter{theorem}{0}
        \setcounter{corollary}{0}
        \setcounter{definition}{0}
        \setcounter{equation}{0}
        \renewcommand{\thefigure}{\Alph{appendixc}.\arabic{figure}}
        \renewcommand{\thetable}{\Alph{appendixc}.\arabic{table}}
        \renewcommand{\theappendixc}{\Alph{appendixc}}
        \renewcommand{\thelemma}{\Alph{appendixc}.\arabic{lemma}}
        \renewcommand{\thetheorem}{\Alph{appendixc}.\arabic{theorem}}
        \renewcommand{\thedefinition}{\Alph{appendixc}.\arabic{definition}}
        \renewcommand{\thecorollary}{\Alph{appendixc}.\arabic{corollary}}
        \renewcommand{\theequation}{\Alph{appendixc}.\arabic{equation}}
        \noindent{\tenbf Appendix \theappendixc #1}\par\vspace{5pt}}
\newcommand{\subappendix}[1] {\vspace{12pt}
        \refstepcounter{subappendixc}
        \noindent{\bf Appendix \thesubappendixc. {\kern1pt \bfit #1}}
	\par\vspace{5pt}}
\newcommand{\subsubappendix}[1] {\vspace{12pt}
        \refstepcounter{subsubappendixc}
        \noindent{\rm Appendix \thesubsubappendixc. {\kern1pt \tenit #1}}
	\par\vspace{5pt}}

\newcommand{\textlineskip}{\baselineskip=13pt}
\newcommand{\smalllineskip}{\baselineskip=10pt}

\def\eightcirc{
\begin{picture}(0,0)
\put(4.4,1.8){\circle{6.5}}
\end{picture}}
\def\eightcopyright{\eightcirc\kern2.7pt\hbox{\eightrm c}}

\def\abstracts#1#2#3{{
	\centering{\begin{minipage}{4.5in}\baselineskip=10pt\footnotesize
	\parindent=0pt #1\par 
	\parindent=15pt #2\par
	\parindent=15pt #3
	\end{minipage}}\par}} 

\renewenvironment{thebibliography}[1]
	{\frenchspacing
	 \ninerm\baselineskip=11pt
	 \begin{list}{\arabic{enumi}.}
	{\usecounter{enumi}\setlength{\parsep}{0pt}
	 \setlength{\leftmargin 12.7pt}{\rightmargin 0pt} 
	 \setlength{\itemsep}{0pt} \settowidth
	{\labelwidth}{#1.}\sloppy}}{\end{list}}

\newcounter{itemlistc}
\newcounter{romanlistc}
\newcounter{alphlistc}
\newcounter{arabiclistc}

\newcommand{\fcaption}[1]{
        \refstepcounter{figure}
        \setbox\@tempboxa = \hbox{\footnotesize Fig.~\thefigure. #1}
        \ifdim \wd\@tempboxa > 5in
           {\begin{center}
        \parbox{5in}{\footnotesize\smalllineskip Fig.~\thefigure. #1}
            \end{center}}
        \else
             {\begin{center}
             {\footnotesize Fig.~\thefigure. #1}
              \end{center}}
        \fi}

\newcommand{\tcaption}[1]{
        \refstepcounter{table}
        \setbox\@tempboxa = \hbox{\footnotesize Table~\thetable. #1}
        \ifdim \wd\@tempboxa > 5in
           {\begin{center}
        \parbox{5in}{\footnotesize\smalllineskip Table~\thetable. #1}
            \end{center}}
        \else
             {\begin{center}
             {\footnotesize Table~\thetable. #1}
              \end{center}}
        \fi}

\def\@citex[#1]#2{\if@filesw\immediate\write\@auxout
	{\string\citation{#2}}\fi
\def\@citea{}\@cite{\@for\@citeb:=#2\do
	{\@citea\def\@citea{,}\@ifundefined
	{b@\@citeb}{{\bf ?}\@warning
	{Citation `\@citeb' on page \thepage \space undefined}}
	{\csname b@\@citeb\endcsname}}}{#1}}

\newif\if@cghi
\def\cite{\@cghitrue\@ifnextchar [{\@tempswatrue
	\@citex}{\@tempswafalse\@citex[]}}
\def\citelow{\@cghifalse\@ifnextchar [{\@tempswatrue
	\@citex}{\@tempswafalse\@citex[]}}
\def\@cite#1#2{{$\null^{#1}$\if@tempswa\typeout
	{IJCGA warning: optional citation argument 
	ignored: `#2'} \fi}}

\def\pmb#1{\setbox0=\hbox{#1}
	\kern-.025em\copy0\kern-\wd0
	\kern.05em\copy0\kern-\wd0
	\kern-.025em\raise.0433em\box0}

\def\fnm#1{$^{\mbox{\scriptsize #1}}$}
\def\fnt#1#2{\footnotetext{\kern-.3em
	{$^{\mbox{\scriptsize #1}}$}{#2}}}

\def\fpage#1{\begingroup
\voffset=.3in
\thispagestyle{empty}\begin{table}[b]\centerline{\footnotesize #1}
	\end{table}\endgroup}

\font\tenrm=cmr10
\font\tenit=cmti10 
\font\tenbf=cmbx10
\font\bfit=cmbxti10 at 10pt
\font\ninerm=cmr9

\font\eightrm=cmr8

\def\qed{\hbox{${\vcenter{\vbox{			
   \hrule height 0.4pt\hbox{\vrule width 0.4pt height 6pt
   \kern5pt\vrule width 0.4pt}\hrule height 0.4pt}}}$}}

\renewcommand{\thefootnote}{\fnsymbol{footnote}}	
\begin{document}

\def\case#1#2{\textstyle{#1\over#2}}

\fpage{1}

\begin{flushright}
JHU--TIPAC--96012\\
hep-ph/9609208\\
August, 1996
\end{flushright}

\vspace{1cm}

\centerline{\bf NEW PREDICTIONS FOR}
\vspace*{0.035truein}
\centerline{\bf CHARMED AND BOTTOM BARYONS\footnote{To appear in the proceedings of the Quarkonium Physics Workshop, University of Illinois, Chicago, June 13--15, 1996.}}
\vspace*{0.37truein}
\centerline{\footnotesize ADAM F.~FALK}
\vspace*{0.015truein}
\centerline{\footnotesize\it Department of Physics and Astronomy, The Johns Hopkins University}
\baselineskip=10pt
\centerline{\footnotesize\it 3400 North Charles Street, Baltimore, Maryland U.S.A.\ 21218}

\vspace*{0.21truein}
\abstracts{I review the recent proposal that there are new isotriplet heavy baryons with masses approximately 2380~MeV and 5760~MeV.  This prediction follows from the application of heavy spin-flavor and light $SU(3)$ symmetries to the observed charmed and bottom baryon states.  It also entails assumptions about the spin and parity quantum numbers of the observed states which are different than is commonly supposed.  The discovery of such states would imply that the nonrelativistic constituent quark model is a poor predictor of heavy baryon spectroscopy.  I update the analysis in light of new data which have become available.}{}{}

\textlineskip
\vspace*{12pt}
\textheight=7.8truein
\setcounter{footnote}{0}
\renewcommand{\thefootnote}{\alph{footnote}}

Many heavy hadrons containing a single charm or bottom quark have been identified in recent years.  While the masses of these particles are usually measured as part of the discovery process, other quantum numbers such as the spin and parity often prove more elusive.  With sufficient data samples such properties can be extracted by studying angular distributions of the particle decays, but these are available only for the lightest and most abundant species.  For others, one typically relies on predictions of models, such as the constituent quark model, to assign quantum numbers to new states.  

This is particularly the case for excited heavy baryons, for which data sets are typically an order of magnitude smaller than for heavy mesons.  In this case, the spin and parity quantum numbers of {\it no\/} states have been measured directly.\fnm{a}\fnt{a}{The decay of the $\Lambda_c$ is known to be consistent with the quark model prediction for its spin, $J=\case12$.\cite{PDG}}  While it is not unreasonable to use quark models as a guide for assigning quantum numbers to heavy hadrons, until recently there has been no test of this approach.  Such a test has now been proposed for the heavy baryons,\cite{Falk} with result that support for the conventional assignments is ambiguous.

The analysis exploits the fact that the bottom and charmed hadrons fall
into representations of heavy quark spin-flavor $SU(4)$ and light flavor
$SU(3)$ symmetries, up to heavy quark corrections of order
$\Lambda_{\rm QCD}/2m_Q$ and $SU(3)$ corrections of order $m_q/\Lambda_\chi$.
Enough states have now been discovered to
make possible detailed tests of the relations implied by the symmetries.
In the heavy meson sector, these predictions are known to work well
for the ground states and the lowest $P$-wave excitations.\cite{FaMe95}
Not only the spectroscopy, but the widths and even the decay angular
distributions are consistent with a simultaneous heavy quark and chiral
$SU(3)$ expansion.  Hence one is tempted to hope that the symmetry
predictions for heavy baryons are also well satisfied. However, in contrast to
the mesons,
for the baryons with the conventional quantum number assignments there are certain symmetry relations which appear to be badly
violated, while others appear to work well.

A variety of resolutions of this situation are available.
First, it is possible that the reported data, with their reported errors, are simply wrong.  While there may be very good reasons to suspect that this is true, this scenario is explicitly outside the scope of this analysis. I will attempt to explain the data as they are currently given in the literature. Second, it is possible that the symmetry breaking corrections
simply are larger than expected.  However such an explanation would
offer no insight into why some relations behave better than others.  Instead, I will propose that the problem is that the conventional
assignment of quantum numbers to the observed charm and bottom baryons may not be
correct.\cite{Falk} In fact, one can satisfy all the symmetry relations at the
expected level by assigning
new quantum numbers to the known resonances.  An exciting consequence is the
existence of additional light excitations which only decay radiatively or weakly.  Such
states are not presently ruled out, and this prediction presents a well defined and conclusive test of the proposal.

I begin with a review of baryon spectroscopy in the heavy quark limit,
$m_c,m_b\to\infty$.  In this limit, heavy quark pair production and
chromomagnetic interactions are suppressed, so the angular momentum and
flavor quantum numbers of the light degrees of freedom become good
quantum numbers.  I will refer to these light degrees of freedom as a
``diquark''; in doing so, I assume nothing about their properties other than
that they carry certain spin and flavor quantum numbers.  For simplicity, I
will also restrict myself for the moment to heavy charm baryons, since the
enumeration of states for bottom baryons is analogous.

In the quark model, the lightest diquark has isospin $I=0$, total spin
$s_\ell=0$ and orbital angular momentum $L_\ell=0$.  With diquark spin-parity
$J_\ell^P=0^+$, this leads to the heavy baryon $\Lambda_c$,
with total $J^P=\case12^+$.  The strange analogue of the
$\Lambda_c$ is the $\Xi_c$, with $I=\case12$.  Because of Fermi statistics,
there is no doubly strange state with $s_\ell=0$.
There is a nearby excitation of the $\Lambda_c$, in which the diquark is in the
same
orbital state, but with $I=s_\ell=1$.  This leads
to a doublet of heavy baryons consisting of the $\Sigma_c$, with
$J^P=\case12^+$, and the $\Sigma^*_c$, with $J^P=\case32^+$.  As with all heavy
doublets, the chromomagnetic hyperfine splitting between these states is of
order $\Lambda_{\rm QCD}^2/m_c$.  The strange analogues of the
$\Sigma_c$ and $\Sigma^*_c$ are respectively the $\Xi'_c$ and $\Xi^*_c$, and
there are also the doubly strange states $\Omega_c$ and $\Omega^*_c$.

The diquark may be excited further by adding a unit of orbital angular
momentum,
$L_\ell=1$.  More precisely, this is true in the constituent quark model, which
guides our intuition that resonances with these
quantum numbers might be close by.  When $I=s_\ell=0$, the excited diquark has
total spin-parity $J_\ell^P=1^-$, and the heavy baryon states are the
$\Lambda_c^*(\case12)$ and the $\Lambda_c^*(\case32)$.  When $I=s_\ell=1$, one
finds diquarks with $J_\ell^P=0^-$, $1^-$ and $2^-$, leading to the odd parity
heavy baryons $\Sigma^*_{c0}$, $\Sigma^*_{c1}(\case12,\case32)$ and
$\Sigma^*_{c2}(\case32,\case52)$.  There are also excited $\Xi_c$ and
$\Omega_c$ baryons.  The spectroscopy and decays of the charm baryons is summarized in Table~\ref{hqstates}.  Multiple decay channels are listed where the dominant decay mode depends on the masses of the states.

\begin{table}[htbp]
  \tcaption{Charm baryon states in the heavy quark limit.  Here $s_\ell$,
$L_\ell$ and $J^P_\ell$ refer respectively to the spin, orbital angular
momentum, and total spin-parity of the light diquark, while $I$ is isospin and
$S$ strangeness.  The given decay channel is the one which is expected to be
dominant, if kinematically allowed.  The enumeration of the bottom baryon
states is analogous.}
\centerline{\footnotesize\smalllineskip
  \begin{tabular}{llllllrl}\\
  Name&$J^P$&$s_\ell$&$L_\ell$&$J^P_\ell$&$I$&$S$&Decay\\ 
  \hline 
  $\Lambda_c$&$\case12^+$&0&0&$0^+$&0&0&weak\\
  $\Sigma_c$&$\case12^+$&1&0&$1^+$&1&0& 
  $\Lambda_c\pi$, $\Lambda_c\gamma$, weak\\
  $\Sigma^*_c$&$\case32^+$&1&0&$1^+$&1&0&$\Lambda_c\pi$\\
  $\Xi_c$&$\case12^+$&0&0&$0^+$&$\case12$&$-1$&weak\\
  $\Xi'_c$&$\case12^+$&1&0&$1^+$&$\case12$&$-1$&$\Xi_c\gamma$, $\Xi_c\pi$\\
  $\Xi^*_c$&$\case32^+$&1&0&$1^+$&$\case12$&$-1$&$\Xi_c\pi$\\
  $\Omega_c$&$\case12^+$&1&0&$1^+$&0&$-2$&weak\\
  $\Omega^*_c$&$\case32^+$&1&0&$1^+$&0&$-2$&$\Omega_c\gamma$\\
  $\Lambda_c^*(\case12)$&$\case12^-$&0&1&$1^-$&0&0&
     $\Sigma_c\pi$, $\Lambda_c\pi\pi$\\
  $\Lambda_c^*(\case32)$&$\case32^-$&0&1&$1^-$&0&0&
     $\Sigma^*_c\pi$, $\Lambda_c\pi\pi$\\
  $\Sigma^*_{c0}$&$\case12^-$&1&1&$0^-$&1&0&$\Lambda_c\pi$\\
  $\Sigma^*_{c1}(\case12,\case32)$&$\case12^-$, $\case32^-$
     &1&1&$1^-$&1&0&$\Lambda_c\pi$\\
  $\Sigma^*_{c2}(\case32,\case52)$&$\case32^-$, $\case52^-$
     &1&1&$2^-$&1&0&$\Lambda_c\pi$\\
  \end{tabular}}
  \label{hqstates}
\end{table}

The masses of these states satisfy a number of heavy quark and
$SU(3)$ symmetry relations.  There are three independent constraints which
relate the bottom and charm systems,
\begin{eqnarray}
   \Lambda_b-\Lambda_c &=&\overline B-
   \overline D=3340\,{\rm MeV}\,,\label{hqrel1}\\
   \overline\Sigma_b-\Lambda_b &=&
   \overline\Sigma_c-\Lambda_c\,,\label{hqrel2}\\
   {\Sigma^*_b-\Sigma_b\over\Sigma^*_c-\Sigma_c} &=&
   {B^*-B\over D^*-D}=0.33\,,\label{hqrel3}
\end{eqnarray}
where in (\ref{hqrel1}) and (\ref{hqrel3}) I have inserted the isospin averaged
heavy
meson masses.\cite{PDG}  Here the states stand for their masses, and a bar
over a state denotes
the spin average over the heavy multiplet of which it is a part.  This spin
average, which cancels the hyperfine interaction between the heavy quark and
the collective light degrees of freedom, takes the form $(D+3D^*)/4$ for the
ground state heavy mesons and $(\Sigma_c+2\Sigma_c^*)/3$ for the
spin-$(\case12,\case32)$ heavy baryon doublets.  The hyperfine
relation~(\ref{hqrel3}) is often written in terms of the ratio
$m_c/m_b$, to which each side is equal, but I prefer a form in which the quark
masses are not introduced explicitly.  The corrections to~(\ref{hqrel1})
and~(\ref{hqrel2}) are expected to be of order
$\Lambda_{\rm QCD}^2(1/2m_c-1/2m_b)\sim50\,$MeV.  The corrections to
(\ref{hqrel3}) could be at the level of~25\%.

The light flavor $SU(3)$ relations are trivial in the exact symmetry limit,
where, for example, $\Sigma_c=\Xi'_c=\Omega_c$.  In this form, of course, they
are also badly violated.  If one includes the corrections linear in
$m_s$, one finds three independent ``equal spacing rules'' for states within the
charm
(or bottom) system,\cite{Savage}
\begin{eqnarray}
   \Omega_c-\Xi'_c &=& \Xi'_c-\Sigma_c\,,\label{su3rel1}\\
   \Omega^*_c-\Xi^*_c &=& \Xi^*_c-\Sigma^*_c\,,\label{su3rel2}\\
   \Sigma^*_c-\Sigma_c &=& \Xi^*_c-\Xi'_c
   = \Omega^*_c-\Omega_c\,.\label{su3rel3}
\end{eqnarray}
Here I neglect isospin violation and electromagnetic effects. The
chiral corrections to the  relations~(\ref{su3rel1})--(\ref{su3rel3})
are expected to be small.\cite{Savage}  There is also a fourth $SU(3)$ relation,
\begin{equation}
   \overline\Sigma_c-\Lambda_c =
   \overline\Xi\vphantom{\Xi}^*_c-\Xi_c\,,\label{su3rel4}
\end{equation}
which is not on the same footing as the others, since it
relates states in two {\it different\/} $SU(3)$ multiplets.\fnm{b}\fnt{b}{This point has recently been emphasized in the literature.\cite{Jenkins,Lebed}}  The leading
corrections to it are, in principle, of order $m_s$, and cannot be
calculated.  However, one's intuition from the quark model is that this
relation should
be reasonably well satisfied, and indeed the counterparts in the charmed meson
sector, such
as $D_{s1}-D_s=D_1-D$, work to within 10~MeV.  In fact all of the heavy quark
and $SU(3)$ relations for the charm and bottom mesons work
beautifully.\cite{FaMe95}

So far, a dozen charm and bottom baryon states have been discovered. I list
them, along with their masses and observed decays, in Table~\ref{baryons}.
However, the names conventionally given to the strongly decaying states imply
certain assumptions
about their quantum numbers and properties.  Since it is precisely these
assumptions which I want to question, I instead identify the observed
resonances
by the modified names listed in the first column of Table~\ref{baryons}. For
simplicity, I have averaged over
isospin multiplets, since isospin breaking is small and not at issue here.

\begin{table}[htbp]
  \tcaption{The observed heavy baryon states, with their conventional and
proposed identities.  Isospin multiplets have been averaged over.  Experimental
errors ($\pm {\rm stat.}\pm {\rm sys.})$ are included where
significant; where they are small, statistical and systematic errors have, for
simplicity, been added in quadrature.  The approximate
masses of the proposed new states are given in parentheses.  (The mass of the $\Xi_{c1}$ is estimated from the plots presented by WA89.  Only
one of the two isospin states has been observed.)}
\centerline{\footnotesize\smalllineskip
  \begin{tabular}{llllll}\\
  State&Mass (MeV)&Ref.&Decay Channel&Conventional&
    Proposed \\ 
  \hline
  $\Lambda_c$&$2285\pm1$&\cite{PDG}&weak&$\Lambda_c$&$\Lambda_c$\\
  &(2380)&&weak&absent&$\Sigma_c^{0,++}$\\
  &(2380)&&$\Lambda_c+\gamma$&absent&$\Sigma_c^+$\\
  $\Sigma_{c1}$&$2453\pm1$&\cite{PDG}&$\Lambda_c+\pi$&$\Sigma_c$&
    $\Sigma^*_c$\\
  $\Sigma_{c2}$&$2519\pm2$&\cite{CLEO96}&$\Lambda_c+\pi$&
    $\Sigma^*_c$&$\Sigma^*_{c0}$\ (?)\\
  $\Xi_c$&$2468\pm2$&\cite{PDG}&weak&$\Xi_c$&$\Xi_c$\\
  $\Xi_{c1}$&$2563\pm15$\ (?)&\cite{WA89}
    &$\Xi_c+\gamma$&$\Xi'_c$&$\Xi'_c$\\
  $\Xi_{c2}$&$2644\pm2$&\cite{CLEO95}&$\Xi_c+\pi$&$\Xi^*_c$&$\Xi^*_c$\\
  $\Omega_c$&$2700\pm3$&\cite{E687}&weak&$\Omega_c$&$\Omega_c$\\ \smallskip
  $\Lambda^*_{c1}$&$2593\pm1$&\cite{PDG,CLEO94}
    &$\Sigma_{c1}+\pi\to\Lambda_c+2\pi$&$\Lambda_c^*(\case12)$&
    $\Lambda_c^*(\case32)$\\ \smallskip
  $\Lambda^*_{c2}$&$2627\pm1$&\cite{CLEO94}&$\Lambda_c+\pi+\pi$&
    $\Lambda_c^*(\case32)$&$\Lambda_c^*(\case12)$\\
  $\Lambda_b$&$5623\pm5\pm4$&\cite{PDG,CDF96}&weak&$\Lambda_b$&$\Lambda_b$\\
  &(5760)&&weak&absent&$\Sigma_b^\pm$\\
  &(5760)&&$\Lambda_b+\gamma$&absent&$\Sigma_b^0$\\
  $\Sigma_{b1}$&$5796\pm3\pm5$&\cite{DELPHI}&$\Lambda_b+\pi$&
    $\Sigma_b$&$\Sigma^*_b$\\
  $\Sigma_{b2}$&$5852\pm3\pm5$&\cite{DELPHI}&$\Lambda_b+\pi$&
    $\Sigma^*_b$&$\Sigma^*_{b0}$\ (?)\\
  \end{tabular}}
   \label{baryons}
\end{table}

The conventional identities of the observed heavy
baryons are given in the fourth column of Table~\ref{baryons}.  How well do the
predictions of heavy quark and $SU(3)$ symmetry fare?  The heavy quark
constraints~(\ref{hqrel1}) and~(\ref{hqrel2}) are both satisfied to within
$10\,$MeV.  However, the hyperfine relation (\ref{hqrel3}) is in
serious trouble.  One finds
$(\Sigma_b^*-\Sigma_b)/(\Sigma_c^*-\Sigma_c)\approx0.84\pm0.14$, too
large by more than a factor of two!  To be conservative, I have ignored the correlation
between the errors on the $\Sigma_b$ and the $\Sigma^*_b$, hence
overestimating the total uncertainty.  It is clear that to take these
data seriously is to identify a crisis for the application of heavy quark
symmetry to the charm and bottom baryons.  By the same token, this crisis rests {\it entirely} on the reliability of the DELPHI measurement\cite{DELPHI} of these states.  Were these data to be called into question, any problem with the heavy quark predictions would disappear.

The situation is also not perfect for the $SU(3)$ relations.  The first equal
spacing rule (\ref{su3rel1}), with the well measured masses of the
$\Sigma_c$ and the $\Omega_c$, yields the prediction $\Xi'_c=2577\,$MeV,
somewhat large but probably within the experimental error.  The second rule
(\ref{su3rel2}) cannot be tested, as the $\Omega^*_c$ state has not yet been
found.  Inserting the measured $\Sigma_c$, $\Sigma^*_c$ and $\Xi_c^*$ masses,
the
third rule~(\ref{su3rel3}) may be rearranged to yield the prediction
$\Xi'_c=2578\,$MeV, again,
reasonably consistent with experiment.  (In fact the precise agreement of these two sum rules might lead one to believe that, when confirmed, the mass of the $\Xi_c'$ will be somewhat lower than its present central value.)  By contrast, the final $SU(3)$ relation~(\ref{su3rel4}) fails by approximately 60~MeV, almost an order of magnitude worse than for the charmed mesons.  Even though this relation is not on the same footing as the others, and its failure is not as significant as the failure of the heavy quark relation (\ref{hqrel3}), such a discrepancy is still surprising and disappointing.

What are we to make of this situation, in which one heavy quark and one
$SU(3)$ relation fail so badly?  If we accept the
quoted experimental errors, perhaps we must also accept that there are
large corrections, that somehow these important symmetries are inapplicable to
heavy baryons.  However, with their striking success in the heavy meson
sector, {\it especially for spectroscopy\/}, it is tempting to look for a new
point of view from which the symmetry predictions are better behaved.

One possibility is to give up the conventional assignments of quantum numbers to the observed heavy baryons.  Since there is no direct evidence for these assignments, they should be chosen to provide the best fit between experiment and theory.  I propose to reinterpret the experimental data under the theoretically motivated constraint that the heavy quark and $SU(3)$ symmetries be imposed explicitly.  Then if we
identify, as before, the observed $\Xi_{c1}$ with the
$\Xi'_c$ state, the $SU(3)$ relations lead to the novel mass
prediction $\Sigma_c\approx2380\,$MeV!  If so, the $\Sigma_c$ cannot be
identified with the observed $\Sigma_{c1}$; in fact, it can be identified with
no resonance yet to have been reported.  However, since at this mass the
$\Sigma_c$ cannot decay in the strong channel $\Sigma_c\to\Lambda_c+\pi$, it
is possible that it exists but so far has been overlooked.

The observed $\Sigma_{c1}$ is now identified as the $\Sigma^*_c$.  In the
bottom baryons, there is a similar reassignment:  the $\Sigma_b$ is now
assumed to be below $\Lambda_b+\pi$ threshold,
while the $\Sigma_{b1}$ is identified as the $\Sigma^*_b$.  As for the
observed $\Sigma_{c2}$ and $\Sigma_{b2}$, they are possibly $I=1$,
$L_\ell=1$ excitations, such as the $\Sigma^*_{c(0,1,2)}$.  While one might
naively estimate that the masses of these states should be larger than those
of the $\Lambda^*_c(\case12)$ and $\Lambda^*_c(\case32)$, a substantial
spin-orbit coupling
could lower the mass of the state $\Sigma^*_{c0}$ by of order $200\,$MeV.
Hence I tentatively identify the observed $\Sigma_{c2}$ and $\Sigma_{b2}$
respectively as the $\Sigma^*_{c0}$ and $\Sigma^*_{b0}$.

The poorly behaved symmetry relations improve dramatically in this
scenario.  For example, let us take the masses of the new states to be
$\Sigma_c=2380\,$MeV and $\Sigma_b=5760\,$MeV.  Then the hyperfine
splitting ratio (\ref{hqrel3}) improves to
$(\Sigma_b^*-\Sigma_b)/(\Sigma_c^*-\Sigma_c)=0.49$, and the $SU(3)$
relation (\ref{su3rel4}) between the $s_\ell=0$ and $s_\ell=1$ states is
satisfied
to within $5\,$MeV.  The heavy quark relation~(\ref{hqrel1}) is unaffected,
while the constraint~(\ref{hqrel2}) for the
$\overline\Sigma_Q$ excitation energy is satisfied to within $20\,$MeV, which
is quite reasonable.  Only the
$SU(3)$ equal spacing rules~(\ref{su3rel1}) and~(\ref{su3rel3}) suffer
from the
change.  Taken, as before, as a prediction for the mass of the $\Xi_c'$, the
former relation now fails by
$23\,$MeV.  The latter now fails by $8\,$MeV, but the discrepancies are in
{\it opposite\/} directions, and the two relations cannot be satisfied
simultaneously by shifting the mass of the $\Xi'_c$.  With these new
assignments, intrinsic
$SU(3)$ violating corrections of the order of $15\,$MeV seem to be unavoidable.  In this context, a confirmation of the $\Xi_c'$ state is very important.  If the mass were to be remeasured to be approximately 2578~MeV, then $SU(3)$ violation under the conventional assignments would be extremely small and we might be more disinclined to relinquish them.

Still, with respect to the symmetry predictions as a whole, the new scenario is a dramatic improvement over the old.  The heavy quark and $SU(3)$ flavor
symmetries have been resurrected.  We can improve the agreement further if we
allow the
measured masses to vary within their reported $1\sigma$ errors.  One set of
allowed
masses is
$\Sigma_c=2375\,$MeV, $\Sigma^*_c=2453\,$MeV, $\Xi'_c = 2553\,$MeV,
$\Xi^*_c=2644\,$MeV, $\Sigma_b=5760\,$MeV, and $\Sigma^*_b=5790\,$MeV.  For
this
choice, the $SU(3)$ relations (\ref{su3rel1}) and (\ref{su3rel3}) (taken as
predictions for the $\Xi'_c$ mass) and (\ref{su3rel4}) are satisfied to within
$15\,$MeV, $13\,$MeV and $4\,$MeV, respectively.  The hyperfine ratio
(\ref{hqrel3}) is $(\Sigma_b^*-\Sigma_b)/(\Sigma_c^*-\Sigma_c)=0.38$, and
$\overline\Sigma_b-\Lambda_b$ is equal to
$\overline\Sigma_c-\Lambda_c$ to within $15\,$MeV.  This is better
agreement with the symmetries than we even have a right to
expect.

Of course, certain problems do remain.  First, and by far the most important, is that
while the new states $\Sigma_c$ (and $\Sigma_b$) have not been
ruled
out, neither have they yet been identified.  These states transform in an isotriplet of $SU(3)$.  One of them, the $\Sigma_c^+$, can decay radiatively in the channel $\Sigma_c^+\to\Lambda_c+\gamma$.  The others, $\Sigma_c^0$ and $\Sigma_c^{++}$, must decay weakly, for example via $\Sigma_c\to\Sigma+\pi$ or $\Sigma_c\to p+K_S+\pi$.  The decays of the new $\Sigma_b$ would be similar.  In the end, their discovery or the
absence thereof will be the defining test of this proposal.

Second, the identification
of
the $\Sigma_{b2}$ as the $\Sigma^*_{b0}$ state, with $s_\ell=L_\ell=1$ and
$J^P_\ell=0^-$, is not entirely satisfactory.  The DELPHI analysis\cite{DELPHI} of the masses, production
and decay properties of the $\Sigma_{b1}$ and $\Sigma_{b2}$ explains in an
elegant
and nontrivial manner the surprisingly low observed polarization of
$\Lambda_b$'s produced at the $Z^0$.\cite{ALEPH,FaPe94}  The analysis was
predicated, of course, on the
conventional assignment of quantum numbers; now this nice explanation of
$\Lambda_b$ depolarization is lost.  Worse, while the $S$-wave decay
$\Sigma^*_{b0}\to\Lambda_b+\pi$ must be isotropic, there appears to be a large
anisotropy
in the direction of the pion in $\Sigma_{b2}\to\Lambda_b+\pi$.  The deviation from an isotropic distribution is about $2.5\sigma$.  If this result
is confirmed, the observed $\Sigma_{b2}$ state must be something else, such as
a radial excitation of the $\Sigma_b^*$.  On the other hand, CLEO\cite{CLEO96} has found no anisotropy in the decay $\Sigma_{c2}\to\Lambda_c+\pi$.  Given the confusing and somewhat inconsistent experimental situation, it is critical that the DELPHI results on $\Sigma_{b1}$ and $\Sigma_{b2}$ be confirmed.  Until then, it will continue to be hard to draw definite conclusions from the data.

Finally, it is worth noting that nonrelativistic constituent quark models  typically do not
favor such a light $\Sigma_c$ and $\Sigma_c^*$ as I have suggested here.  (See, for example, recent papers by Lichtenburg\cite{Lich} and Franklin,\cite{Fran} as well as the papers cited by Savage.\cite{Savage})  These models have been useful for predicting hadron masses, and are thus, not unreasonably, quite popular.  However,
despite common misperceptions,\cite{Lich,Fran} they are {\it less\/} general, and make substantially {\it more\/} assumptions, than a treatment based solely on heavy quark and $SU(3)$ symmetry.  A reasonable quark model respects these symmetries in the appropriate limit, as well as parametrizing deviations from the symmetry limit.
Such models therefore cannot be reconciled simultaneously with the heavy quark
limit and with the reported masses of the $\Sigma_b$ and $\Sigma_b^*$.  Hence,
the predictions of this analysis follow experiment in pointing to physics beyond
the constituent quark model.  While the historical usefulness of this model for hadron spectroscopy may deepen one's suspicion of the DELPHI data on $\Sigma_{b1,2}$, such speculation is beyond the scope of this discussion.  To reiterate, I have taken the masses and errors of all states as they have been reported to
date; as they evolve in the future, so, of course, will the theoretical
analysis.

While such issues are important, the smoking gun here is the
prediction of new weakly and radiatively decaying heavy baryon excitations.  If confirmed, this will
be the
most unexpected and striking prediction yet to be obtained from heavy quark
symmetry.  If not, and if the reported data and conventional quantum number assignments are correct, we will have to accept
the failure of heavy spin-flavor
and light $SU(3)$ symmetry to describe the charm and bottom baryon states.

\nonumsection{Acknowledgements}

It is a pleasure to thank the organizers of the workshop, especially Howard Goldberg, Tom Imbo, and Wai-Yee Keung, for a most enjoyable and productive meeting.  That the Bulls lost Games 4 and 5 of the NBA Finals, thereby postponing the victory celebration until after the workshop had ended, was no fault of theirs.  I am grateful to Vivek Sharma and John Yelton for helpful conversations concerning experimental issues.  This work was supported by the National Science Foundation under Grant No.~PHY-9404057 and National Young Investigator Award No.~PHY-9457916, by the Department of Energy under Outstanding Junior Investigator Award No.~DE-FG02-94ER40869, and by the Alfred P.~Sloan Foundation.

\nonumsection{References}

\end{document}